\documentclass[aps,prl,preprint,superscriptaddress,showpacs,endfloats]{revtex4-1}
\usepackage[T1]{fontenc}
\usepackage{amsmath}
\usepackage[latin1]{inputenc}
\usepackage{graphicx}

\usepackage{epsfig}
\usepackage{color}

\newcommand{\mub}{\mu^b}

\begin{document}

\title{Phase stacking diagram of colloidal mixtures under gravity}

\author{Daniel de las Heras}
\affiliation{Theoretische Physik II, Physikalisches Institut, 
  Universit{\"a}t Bayreuth, D-95440 Bayreuth, Germany}
\author{Matthias Schmidt}
\affiliation{Theoretische Physik II, Physikalisches Institut, 
  Universit{\"a}t Bayreuth, D-95440 Bayreuth, Germany}

\date{\today}

\begin{abstract}
The observation of stacks of distinct layers in a colloidal or liquid
mixture in sedimentation-diffusion equilibrium is a striking
consequence of bulk phase separation. Drawing quantitative conclusions
about the phase diagram is, however, very delicate. Here we introduce
the Legendre transform of the chemical potential representation of the
bulk phase diagram to obtain a unique stacking diagram of all possible
stacks under gravity. Simple bulk phase diagrams generically lead to
complex stacking diagrams. We apply the theory to a binary hard core
platelet mixture with only two-phase bulk coexistence, and find that
the stacking diagram contains six types of stacks with up to four
distinct layers. These results can be tested experimentally in
colloidal platelet mixtures. In general, an extended Gibbs phase rule
determines the maximum number of sedimented layers to be
$3+2(n_b-1)+n_i$, where $n_b$ is the number of binodals and $n_i$ is
the number of their inflection points.
\end{abstract}

\maketitle

Adding a second component to a colloidal dispersion generates a wealth
of new phenomena, such as the induction of effective interactions of
the primary colloidal component by adding a depletion agent
\cite{tuinier2003depletion}, the formation of complex ionic colloidal
crystals by adding an oppositely charged component
\cite{leunissen2005ionic}, and the self-assembly of patchy colloidal
mixtures into two interpenetrating networks
\cite{C1SM06948A,varrato2012arrested}. Binary colloidal crystals are
promising candidates for {\em e.g.}\ photonic applications
\cite{vermolen}. Gravity can have a strong effect on colloidal
mixtures because on typical length scales in the lab the gravitational
energy and the thermal energy are comparable. The resulting
observations are often very counter-intuitive, including denser
particles floating on top of a fluid of lighter colloidal spheres
\cite{C2SM26120K}, or the emergence of a floating nematic phase
sandwiched between two isotropic phases in mixtures of colloidal
platelets and spheres \cite{floating}.  The number of distinct layers
in a sample in sedimentation-diffusion equilibrium can be considerably
large: {\em e.g.}~a vertical stack of six different phases was
observed in mixtures containing charged colloidal platelets
\cite{doi:10.1021/la804023b}.  Much attention was devoted to
sedimentation in charged colloidal systems \cite{Philipse}.  Although
the investigation of the bulk phase behaviour of colloidal mixtures is
often based on sedimentation experiments \cite{C1SM06535A}, the
interpretation of the experimental findings, and in particular drawing
quantitative conclusions about the bulk phase diagram, can be a very
subtle issue. Understanding and controlling the stacking sequence of
colloidal mixtures is of potential interest in industrial
applications, such as preventing phase separation in commercial
colloidal products and inducing demixing in order to sort colloidal
particles by size or density \cite{serrano}.  In molecular rather than
colloidal mixtures corresponding effects due to gravity arise on
larger length scales, as {\em e.g.}\ relevant for species segregation with
depth in oil reservoirs \cite{Debenedetti1988,EspositoEtAl2000}.

Here we show for the first time that the phenomenology of all possible
stacking sequences of a colloidal or liquid mixture is directly
related to the bulk phase diagram of the system, without requiring 
further information about the equation of state.  A systematic
stacking diagram follows from the bulk phase diagram by Legendre
transform in a unique way. Here the Legendre transform acts on the
features of the bulk phase diagram, such as the binodal(s), critical
point(s) {\em etc.}, rather than on the thermodynamic
potential(s). The Legendre transform introduces the ratio of the
buoyant masses as a control parameter. In experiments this could be
varied by {\em e.g.}~changing the solvent of a colloidal mixture. We
apply the theory to predict the stacking diagram of a mixture of
colloidal platelets, using density functional theory to first obtain
the bulk phase diagram. The mixture models a binary smectite-gibbsite
platelet system, and hence the results can be tested experimentally.

\section{Theory}

We base our theory on the chemical potential as the central
quantity. In the presence of gravity, one can define a height- and
species-dependent local chemical potential
\cite{hansen06,schmidt04aog,floating}:
\begin{align} 
  \psi_i(z)=\mub_i-m_igz,
  \label{EQlocalchemichalPotential}
\end{align}
where $z$ is the vertical coordinate, $\mub_i$ is the chemical
potential of species $i$, $m_i$ is its buoyant mass and $g$ is the
acceleration due to gravity. Eq.~(\ref{EQlocalchemichalPotential})
immediately implies that the difference in the local chemical
potential between two different heights is experimentally accessible
by measuring the difference in height between the two points.
Eliminating $gz$ from (\ref{EQlocalchemichalPotential}) yields
\begin{align}
  \psi_2(\psi_1)&=a+s\psi_1,\label{EQpath}
\end{align}
where the constants are $a=\mub_2-s\mub_1=$ and
$s=m_2/m_1$. Eq.~(\ref{EQpath}) describes a straight line in the plane
of local chemical potentials $\psi_1,\psi_2$.  The slope $s$ of this
``sedimentation path'' is given by the ratio of the buoyant masses or,
equivalently, by the inverse ratio of the gravitational heights,
$s=\xi_1/\xi_2$, where $\xi_i=k_BT/(m_ig)$ is the gravitational height
of species~$i$; here $k_B$ is the Boltzmann constant and $T$ is
absolute temperature.

Eq.~(\ref{EQpath}) attains great significance when combined with a
local density approximation (LDA) \cite{hansen06}, which applies when
all relevant correlation lengths in the system are small compared to
all $\xi_i$; this is analogous to dividing the system in small
horizontal slabs, which are treated as individual equilibrium systems
\cite{Debenedetti1988,EspositoEtAl2000}.  Then one can assume that the
state of the system at height $z$ is analogous to a bulk state with
chemical potentials $\mu_i$ and that
\begin{align}
  \mu_i=\psi_i(z).
\end{align}
As a consequence, the sedimentation path (\ref{EQpath}) is directly
related to the experimentally observed stacking sequence in the vessel
\cite{schmidt04aog,floating}.  Local phase coexistence between A and B
occurs at height $z_{\rm AB}$, provided that $\psi_i(z_{\rm
  AB})=\mu_{i, \rm AB}$ for both species; here $\mu_{i, \rm AB}$ is
the chemical potential of species $i$ at bulk coexistence between
phases A and B. As a result a (horizontal) AB interface at height
$z_{\rm AB}$ is observed in the sample.

\begin{figure}
\includegraphics[width=0.6\columnwidth]{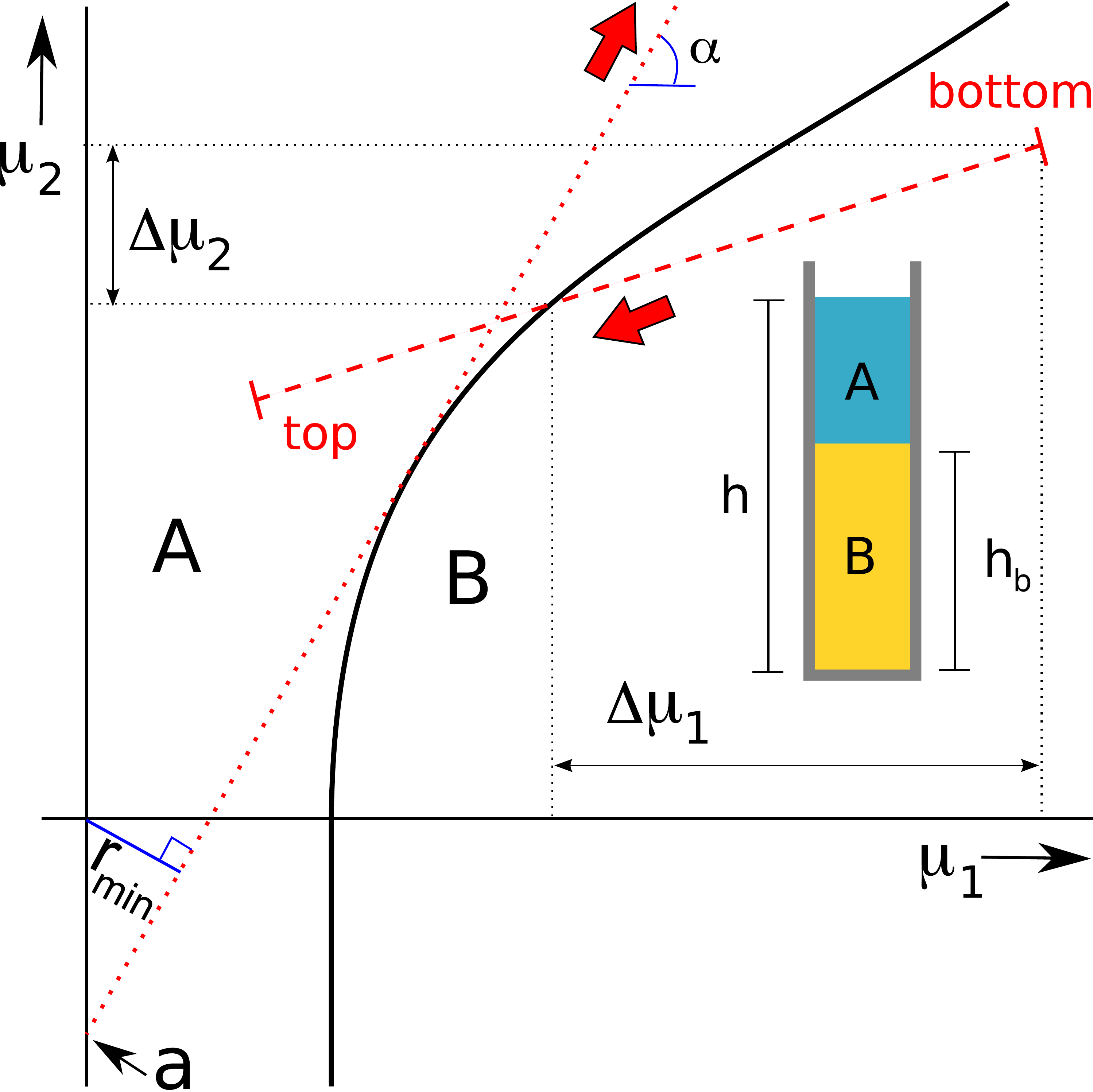}
\caption{ Schematic representation of the bulk phase
  diagram of a colloidal mixture in the plane of chemical potentials
  $\mu_2,\mu_1$. The black-solid line represents the binodal where the
  phases A and B coexist. The red-dashed line represents a
  sedimentation path. Its direction (from bottom to top of the sample)
  is indicated by an arrow. The inset shows the stacking sequence and
  the relative thickness of each sedimentation layer corresponding to
  the sedimentation path. The dotted red line represents a
  sedimentation path tangent to the bulk binodal.}
\label{fig1}
\end{figure}

In order to illustrate these effects, we plot in Fig.~\ref{fig1}
a~schematic bulk phase diagram of a mixture with stable A and B bulk
phases.  We show a sedimentation path that starts in the region where
B is stable, crosses the binodal, and ends in the region of stability
of A. The corresponding stacking sequence consists of bottom B and top
A, which we write as BA. Clearly, the thickness of each sedimentation
layer is proportional to the difference in chemical potentials between
its upper and lower interface,
cf.\ Eq.~(\ref{EQlocalchemichalPotential}).  The difference in
chemical potentials between the crossing of the sedimentation path and
the binodal and the bottom of the sample in Fig.~\ref{fig1} is simply
$\Delta\mu_i=-m_igh_{\rm B}$, with $h_{\rm B}$ the macroscopic
thickness of the bottom sedimentation layer.  Neglecting effects due
to finite sample height (to which we will turn below), the conditions
for a sedimented sample are fully determined by the parameters $a$ and
$s$, which we now treat as new (state) variables. Each pair $a$, $s$
determines uniquely a sedimentation path (\ref{EQpath}). We will in
the following identify three types of boundaries in the $a,s$ plane,
where the sedimentation behaviour changes qualitatively upon
infinitesimal deviation of the parameters. We refer to these
boundaries as: the sedimentation binodal, the terminal line, and the
asymptotic terminal line.

(i) {\em Sedimentation binodal.}
As $T=\rm const$, the Gibbs phase rule leaves one remaining free
thermodynamic variable to parameterize $\mu_{i,\rm AB}$. Choosing this
parameter as $\mu_1$ results in $\mu_{1, \rm AB}=\psi_1$ and
$\mu_{2,\rm AB}(\mu_1)=\psi_2(\psi_1)$, from which one obtains
at coexistence
\begin{align}
  \mu_{2,\rm AB}(\psi_1) &= \psi_2(\psi_1).
  \label{EQcrossing}
\end{align}
The marginal case is obtained when the slope of the sedimentation path
equals the slope of the binodal, which implies that
\begin{align}
  \mu_{2, \rm AB}'(\psi_1) &= \psi_2'(\psi_1),
  \label{EQtangent}
\end{align}
is simultaneously satisfied with (\ref{EQcrossing}); in
(\ref{EQtangent}) the prime denotes the derivative with respect to the
argument. Multiplying (\ref{EQtangent}) by $\psi_1$, subtracting it
from (\ref{EQcrossing}), and observing the structure (\ref{EQpath})
yields
\begin{align}
  a_{\rm AB}(s) = \mu_{2,\rm AB}(\psi_1) - \psi_1 \mu_{2,\rm AB}'(\psi_1),
  \label{EQlegendreTransform}
\end{align}
where $a_{\rm AB}(s)$ is the intercept of the marginal case.
Eq.~(\ref{EQlegendreTransform}) establishes $a_{\rm AB}(s)$ as the
Legendre transform of the chemical potential representation of the
bulk binodal. Hence any path (\ref{EQpath}), where the parameters are
linked via $a=a_{\rm AB}(s)$, is special in that it divides the $a,s$
parameter space into distinct regions. These regions differ in that a
qualitative change in the stacking sequence occur, namely the
emergence of a floating phase ({\em e.g.}~from A to ABA). All paths that
satisfy (\ref{EQlegendreTransform}) are tangent to the binodal (see
Fig.~\ref{fig1}).

(ii) {\em Terminal line.} The systems reaches a critical point, when
locally for both components
\begin{align}
  \psi_{i,\rm crit}(z_{\rm crit}) = \mu_{i,\rm crit},
\end{align}
from which follows, upon using the general expression (\ref{EQpath}),
that
\begin{align}
  a_{\rm crit}(s) = \mu_{2,\rm crit} - s \mu_{1,\rm crit},
  \label{EQcriticalPaths}
\end{align}
which is linear in $s$, implying that a critical point corresponds to
a straight (terminal) line in the $a,s$-space of sedimentation
paths. Indeed the same reasoning can be applied to the triple point
and other ``special'' points in the phase diagram, at which a binodal
ends. In each of these cases the result is analogous to
(\ref{EQcriticalPaths}).  The sedimentation paths that correspond to
(\ref{EQcriticalPaths}), {\em i.e.}~that lie on the terminal line, are
those that cross the special point.

(iii) {\em Asymptotic terminal line.} A bulk binodal that does not
terminate at finite values of $\mu_i$ can either be connected to a
phase transition of one of the pure subsystems or it can represent a
large demixing region at very high chemical potentials. In both cases
the bulk binodal tends to an asymptote with well-defined slope,
\begin{align}
  \mu_{2,\rm AB}/\mu_{1,\rm AB}\to s_\infty,
  \label{EQasymptoticTerminalLine}
\end{align}
and a corresponding (asymptotic terminal) line emerges in the stacking
diagram, which is described by $s(a)=s_\infty=\rm const$, for all
values of $a$.  The paths that are described by
(\ref{EQasymptoticTerminalLine}) are those that are parallel to the
asymptote of the binodal.

\section{Results}

\begin{figure}
\includegraphics[height=0.65\textheight]{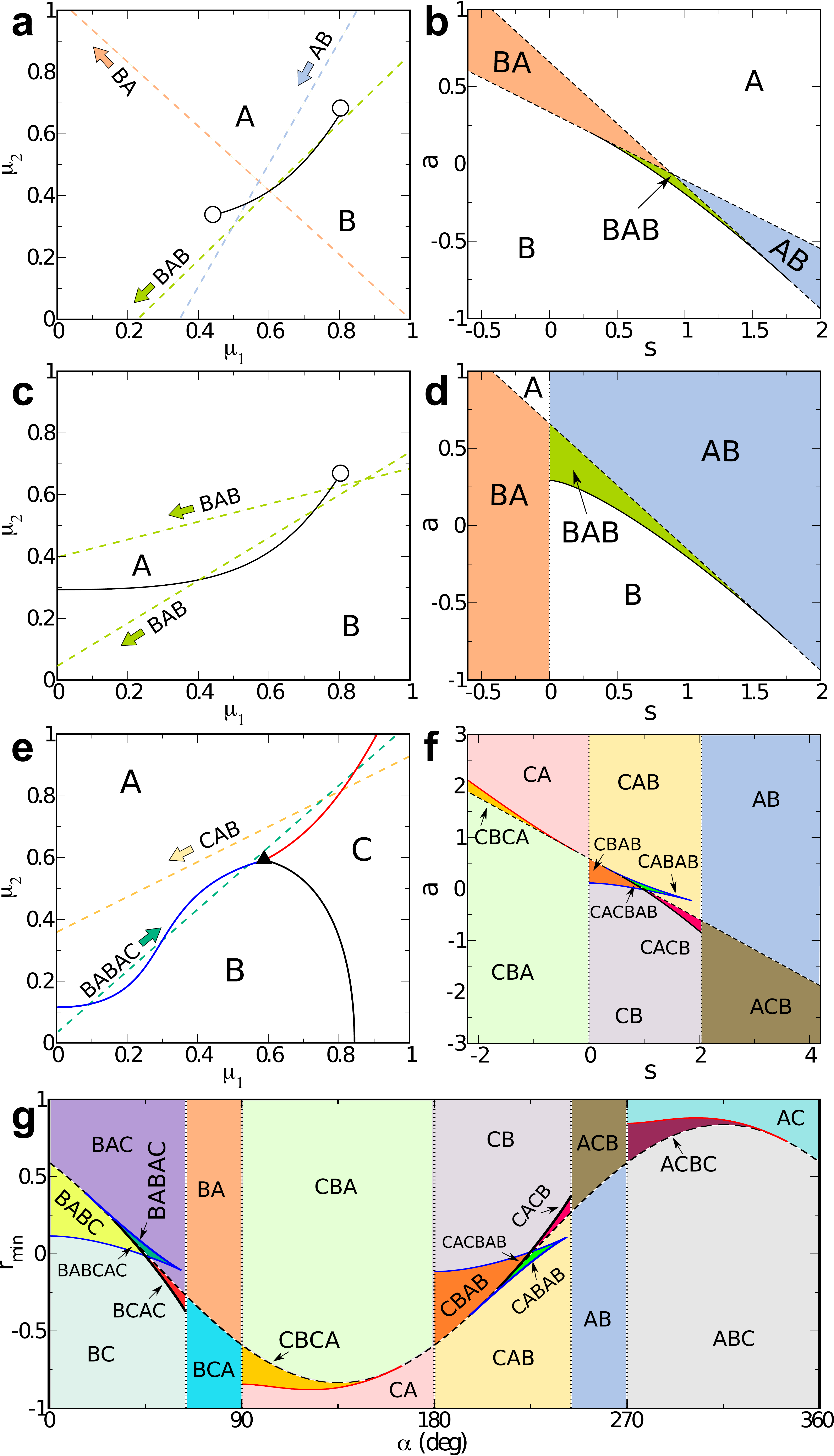}
\caption{Schematic bulk phase diagrams of colloidal
  mixtures in the plane of chemical potentials $\mu_1,\mu_2$ (a,c,e),
  and the corresponding stacking diagrams in the plane of slope $s$
  and intercept $a$ of the sedimentation path (b,d,f); here we set
  $m_1>0$. Chemical potentials are given in arbitrary units.  (g)
  Sedimentation phase diagram of the mixture of panel (e) in the
  $\alpha,r_{\rm min}$ plane. In panels (a,c,e) solid lines represent
  binodals (modelled as B\'ezier curves), empty circles represent
  critical points, dashed lines indicate selected sedimentation paths
  (the arrows gives the direction from bottom to top of the sample,
  labels indicate the corresponding stacking sequences). The triangle
  in (e) represents the triple point. In panels (b,d,f,g) the solid
  lines indicate sedimentation binodals, dashed lines represent
  terminal lines, dotted lines indicate asymptotic terminal
  lines. Different stacking regions are coloured and are labelled by
  their respective stacking sequence.}
\label{fig2}
\end{figure}

We show in Fig.~\ref{fig2} three schematic bulk phase diagrams in the
$\mu_1,\mu_2$ plane (a,c,e) and the corresponding stacking diagrams in
the $a,s$ plane (b,d,f). The simplest example (a) consists of a
binodal that ends at two critical points. This topology corresponds to
a closed immiscibility loop in the pressure-composition or
density-density plane, as predicted, {\em e.g.}, for mixtures of
patchy colloids \cite{C0SM01493A}.  Two sedimentation
paths with the same slope but opposite directions correspond to
stacking sequences with inverse order ({\rm e.g.}, AB and BA); in
order to avoid this ambiguity we only consider the case $m_1>0$. The
corresponding stacking diagram (b) contains one sedimentation binodal
and one terminal line for each of the two critical points. Each
terminal line is tangent to its end of the sedimentation binodal.  The
crossing point, where the two terminal lines intersect, represents the
sedimentation path that connects the two critical points in the bulk
phase diagram. The sedimentation binodal and the terminal lines divide
the $a,s$ plane into five different regions with differing stacking
sequences. Those paths that cross the bulk binodal twice generate the
stacking sequence BAB. This triple stacks occurs although  there is no triple point in
the bulk phase diagram nor are there three different phases. A similar
stacking sequence, a nematic phase sandwiched between two isotropic
phases, has been theoretically predicted and experimentally 
observed very recently \cite{floating} in mixtures of
silica spheres and gibbsite platelets.  In general, two consecutive
phases in the stacking sequence coexist in bulk, but two
non-consecutive phases may or may not coexist in bulk and, as the
example demonstrates, the same phase (B) can reenter the sequence.  An
obvious consequence is that the Gibbs phase rule does not apply to the
maximum number of layers in sedimentation equilibrium.  Instead, the
maximum number of sedimented layers in a mixture under gravity is $3 +
2(n_b - 1) + n_i$, where $n_b$ is the number of bulk binodals and
$n_i$ is the total number of inflection points in all binodals. This
maximal number of layers occurs if a sedimentation path crosses each
binodal $2 + n_{b,i}$ times, with $n_{b,i}$ the number of inflection
points of each binodal. The occurrence of an inflection point
in a binodal has been predicted in colloidal platelet-sphere mixtures \cite{floating}. 
It is also a natural consequence of a binodal that connects two phase transitions 
in the pure components of the mixture. An example is the isotropic-nematic 
transition in binary platelets mixtures, to which we turn below.

We next consider a case (c) where the AB phase transition persists in
the pure system of species 2. The bulk binodal tends asymptotically to
the value of $\mu_2$ at the phase transition of the pure species 2
when $\mu_1\rightarrow-\infty$. In the stacking diagram (d) an
asymptotic terminal line appears: as laid out above, this vertical
line is at a position $s$ that gives the slope of the asymptote of the
bulk binodal.

The complexity of the stacking diagram increases very significantly
with the number of stable phases in the bulk phase diagram. In
Fig.\ref{fig1} (e) we show a still rather simple bulk phase diagram
with three bulk phases A, B and C that coexist at a triple point. The
pure species~1 undergoes a BC phase transition that persists in the
mixture and ends at the triple point. The BC bulk binodal tends
asymptotically to the value of $\mu_1$ at the phase transition of the
monodisperse system of species $1$ when $\mu_2\rightarrow-\infty$
(vertical asymptote). The component $2$ of the mixture undergoes an AB
phase transition. The binodal has a horizontal asymptote and ends at
the triple point.  At chemical potentials above the triple point there
is strong demixing between A and C. The AC binodal has asymptotically
a well-defined slope, for which the general expression holds:
$\left.d\mu_2/d\mu_1\right|_{\text{\rm
    coex}}=\Delta\rho_1/\Delta\rho_2$, with $\Delta\rho_i$ the jump in
density of species $i$ at the phase transition. At very high chemical
potentials both species approach the close packing density and the
slope of the binodal will be constant. The corresponding stacking
diagram, Fig.\ref{fig2} (f), has three sedimentation binodals (one for
each bulk binodal), one terminal line due to the triple point, and two
asymptotic terminal lines. The asymptotic terminal line of the BC bulk
binodal is not visible because its slope is infinite. To overcome this
problem, we show in panel (g) the stacking diagram in the
$\alpha,r_{\rm min}$ plane, where $r_{\rm min}$ is the distance of the
sedimentation path to the origin in the $\mu_1,\mu_2$ plane and $\tan
\alpha=s$, cf.\ Fig.~\ref{fig1}. Here we use the sign convention that
$r_{\rm min}$ is positive (negative) for a clockwise
(counterclockwise) path in order to discriminate between paths with
identical value of $\alpha$, but negative value of $s$. We find six
sedimentation binodals (two for each bulk binodal because in this
plane we plot all possible paths, $0\le\alpha\le2\pi$, lifting any
sign restriction on $m_1$), one terminal line, and six asymptotic
terminal lines at $\alpha=0,0.35\pi,\pi/2,\pi,1.35\pi$, and
$3\pi/2$. Each pair $\alpha$ and $\alpha+\pi$ represents the
asymptotic behaviour (and direction) of a bulk binodal in the stacking
diagram. All intersections between sedimentation binodals and terminal
lines define the boundaries of the sedimented phases.  There are 22
different stacking sequences, which vary in complexity from the simple
AC to the exotic CACBAB. Sedimentation monophases do not occur, as
each path crosses at least one binodal. These examples demonstrate the
extreme richness of the stacking diagram.

\begin{figure}
\includegraphics[width=0.9\columnwidth]{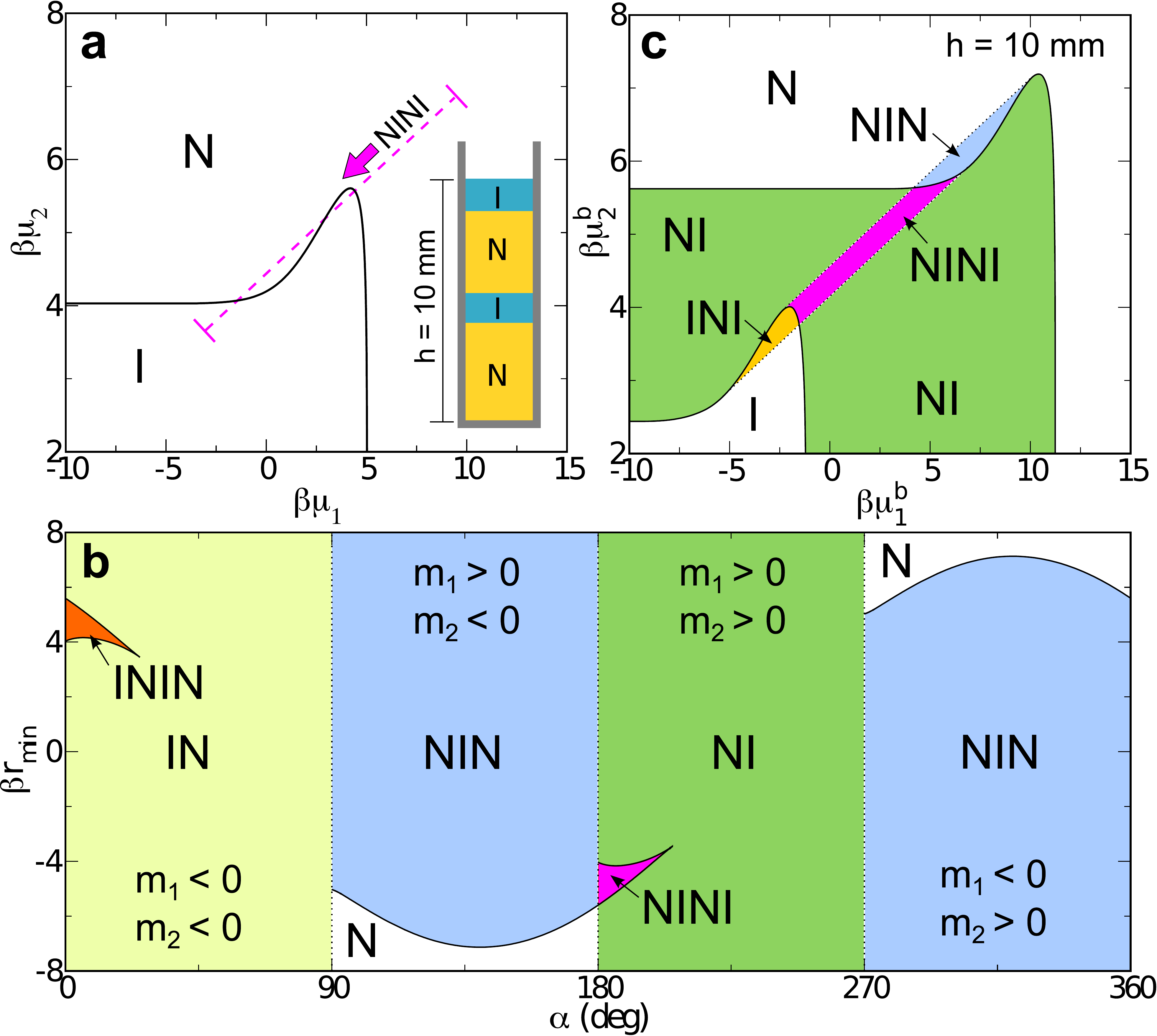}
\caption{Phase diagram and stacking diagram of a binary mixture of colloidal platelets.
  (a) Bulk phase diagram as a function of the
  chemical potentials $\mu_1$ and $\mu_2$ of small
  platelets (species $1$) and big platelets (species $2$) with size
  ratio $R_2/R_1=1.4$. The solid line represents the IN bulk
  binodal. The dashed line indicates a selected sedimentation path for
  a vessel with height $h=10\text{ mm}$. The arrow gives the direction
  from bottom to top of this sample. The inset illustrates the
  corresponding stacking sequence. (b) Sedimentation phase diagram for
  infinite sedimentation paths in the $\alpha,r_{\rm min}$ plane. The
  solid lines represent the sedimentation binodals, and the vertical
  dotted lines indicate the asymptotic terminal lines. (c)
  Sedimentation phase diagram in the plane of (scaled) reference
  chemical potentials $\beta\mub_1, \beta\mub_2$,
  cf.\ Eq.~(\ref{EQlocalchemichalPotential}), for sample height
  $h=10\text{ mm}$, {\em i.e.}~sedimentation paths of finite
  length. The solid lines represent the sedimentation binodals. Each
  colour in (b) and (c) represents a different sedimentation state,
  labeled by its corresponding stacking sequence upon increasing
  height.}
\label{fig3}
\end{figure}

We next apply the theory to a mixture of infinitely thin circular hard
platelets with aspect ratio $R_2/R_1=1.4$, where $R_i$ is the radius
of species $i$. We use a microscopic geometry-based density functional
to calculate the bulk phase diagram. The functional goes beyond the
Onsager limit and has been used previously to analyze the bulk phase
diagram of platelet-platelet mixtures with different aspect ratios
\cite{PhysRevE.81.041401}. The pure fluid of hard platelets undergoes
an isotropic-nematic (IN) phase transition as a function of the
chemical potential.  Further phases with positional order, such as
columnar or crystal phases, are not present in the model due to the
vanishing thickness of the particles \cite{PhysRevE.57.4824}.  The
bulk phase diagram of the binary mixture is depicted in the
$\mu_1,\mu_2$ plane in Fig.~\ref{fig3} (a).  The IN bulk binodal
connects the phase transitions of the pure components.  The chemical
potential of the small platelets, $\mu_1$, decreases when a small
fraction of big platelets is added to the pure fluid of small
platelets (big platelets favour the orientational order of the small
platelets). The behaviour of $\mu_2$ is the opposite: its coexistence
value increases if a tiny fraction of small platelets is added to the
pure system of big platelets. As a result, the bulk binodal, that
connects both limits, has a curvature change and a maximum at
intermediate values of composition.

The stacking diagram of the mixture is plotted in Fig.~\ref{fig3} (b)
in the $\alpha,r_{\rm min}$ plane. There are two sedimentation
binodals and four asymptotic terminal lines at $\alpha=0,\pi/2,\pi$
and $3\pi/2$. We can identify six different sedimentation sequences,
including a floating isotropic phase NIN, a double floating isotropic
NINI and nematic ININ. Such states with four layers arise from the
curvature change of the bulk binodal. The $NINI$ four stack can be
understood as a fractionation between two double $NI$ stacks, the
lower one being rich in heavier platelets, and the upper one being
rich in the lighter platelets. This interpretation applies to the
inverted $ININ$ sequence, which appears in the region of negative
buoyant masses.

Real samples possess a finite height $h$, such that there is only an
interval of the corresponding infinite sedimentation path
accessible. In order to characterize such a state, four variables are
required, which we choose to be the slope $s$ of the path, the two
reference chemical potentials $\mub_1$ and $\mub_2$,
cf.~Eq.~(\ref{EQlocalchemichalPotential}), and the sample height
$h$. We adopt the convention that the origin of the coordinate system
is in the middle of the sample, such that $-h/2 < z< h/2$. As a
consequence at $z=0$ the local chemical potentials $\psi_i(z)=\mub_i$.
We choose $\xi_1=0.805\text{ mm}$ and $\xi_2=3.89\text{ mm}$, which is
compatible with, {\em e.g.}, a colloidal mixture of gibbsite platelets
(species $1$: average diameter $214\text{ nm}$ and thickness $10\text{
  nm}$) and smectite (species $2$: average diameter $300\text{ nm}$
and thickness $1\text{ nm}$) dispersed in water at room temperature
(see {\em e.g.}~\cite{floating} and references therein).  We obtain
$s=\xi_1/\xi_2=0.207$ or $\alpha=192^{\circ}$, and we choose the
vessel height as $h=10\text{ mm}$.  An example of a sedimentation path
and the corresponding sedimentation sequence for these conditions is
shown in Fig.~\ref{fig3} (a).  The stacking diagram for finite height
is shown in Fig.~\ref{fig3} (c).  Due to the finite sample height,
additional sedimentation binodals arise when a sedimentation path
starts or ends at the bulk binodal.  While for infinite sedimentation
paths there are only two sedimentation phases, NI and NINI, for finite
height we find six different stacks I, N, NI, NIN, INI, and
NINI. Hence the main effect of the finite length of the sedimentation
path is that new sedimentation stacks appear. These are formed by
removing layer(s) on top or bottom of the infinite stacking
sequences. Note that the $\mub_1,\mub_2$ plane can be converted to the
plane of average densities by calculating the density profiles and
average density of those paths that form the phase boundaries of the
stacking diagram.

\section{Discussion}

A colloidal mixture under gravity can be represented by a straight line
in the plane of chemical potentials, which we refer to as a
sedimentation path. A crossing between the sedimentation path and a
binodal corresponds to an interface in the sample, which establishes a
direct relation between the path and the observed phase stacking
sequence under gravity.  We have developed a general theory that
relates all possible stacking sequences of a colloidal mixture to its
bulk phase diagram. We have shown how to group the stacking sequences
in a stacking diagram that follows in a unique way from the bulk phase
diagram. The binodals, their ending points, and their asymptotic
behaviour are the three distinct elements that determine the
boundaries between the different sedimentation states in the stacking
diagram. The stacking and the bulk phase diagrams are linked by a well
defined mathematical mapping based on the Legendre transform.

In order to characterize the sedimentation states, a suitable order
parameter is the array of thicknesses, $h_m$, of the individual layers
$m=1,2,\ldots$ in a sedimented stack. When crossing a boundary in the
sedimentation diagram, at least one of the $h_m \to 0$, which implies
a continuous phase transition in the inhomogeneous
systems. Finite-size and surface effects that are beyond our LDA
treatment, will modify and enrich this scenario via wetting at the
upper and lower boundaries and capillary evaporation effects of very
thin floating layers \cite{schmidt04aog}.

The order of the individual stacks is always of decreasing total mass
density, $\rho_m(z)=\sum_i m_i\rho_i(z)$. The condition $0\geq
d\rho_m(z)/dz\equiv \sum_{ij} m_i(\partial \rho_i/\partial
\mu_j)d\psi_j(z)/dz$ follows by imposing local thermodynamic stability
along the sedimentation path, {\em i.e.}~that the matrix of partial
derivatives $\partial \rho_i/\partial \mu_j|_{T,V,\mu_k}$, where $V$ is the
system volume, is positive definite \cite{Reichl98}. This is
consistent with the findings in
\cite{Debenedetti1988,EspositoEtAl2000}, and not in conflict with the
possibility of denser particles floating on top of lighter
ones, as recently reported by Piazza et al~\cite{C2SM26120K}, provided
that the floating phase has lower mass density than the supporting
phase.

Gravity induces very rich phenomenology even for the simplest of
mixtures. For example, the binary hard core platelet mixture that we
investigated has only isotropic and nematic phases stable in bulk. We
found that the stacking diagram contains six different types of
sequences, with up to four distinct layers. The maximum number of
layers in a stacking sequence is not limited by the Gibbs phase rule
for bulk phases. Instead, an extended Gibbs phase rule holds, which
accounts for the maximum number of layers that can appear in
sedimentation-diffusion equilibrium. The relevant parameters are the
number of binodals and the number of their inflection points, rather
than only the number of components as is relevant in bulk.  As a
consequence, extremely rich stacking sequences can be found even for
very simple mixtures. An example is the NINI stack that we predict to
occur in the binary platelet mixture. Mixtures with three binodals can
lead to stacking sequences with seven distinct layers in the simplest
case with no inflection points.

Our results demonstrate that the ratio between the buoyant masses of
both species, $s$, is a key parameter that controls the stacking
sequence of a mixture.  Changing the strength of gravity, {\em
  i.e.}~by using centrifugal forces as in analytic centrifugation,
leaves the slope $s$ invariant. However, if the species are made of
different materials, one can vary $s$ experimentally by changing the
density of the solvent, because
$m_2/m_1=v_2(\rho_{m,2}-\rho_s)/(v_1(\rho_{m,1}-\rho_s))$, where $v_i$
is the particle volume, $\rho_{m,i}$ is the mass density of
species~$i$, and $\rho_s$ is the solvent density. Alternatively, one
could vary $s$ by designing colloids with cores made of different
materials \cite{blaaderen}. While clearly the average density of
colloids in the sample (related to $a$ cf.\ Eq.~(\ref{EQpath})) plays
a major role in determining the stacking sequence, our considerations
for the binary platelet mixture demonstrate that the total height of
the sample is a further relevant parameter that should be carefully
controlled in sedimentation experiments.

Our approach is relevant for computer simulation work, as the
sedimentation-diffusion equilibrium of colloidal mixtures can be
predicted by simulating only the bulk phase behaviour, which can be
less computationally demanding than performing simulations of
inhomogeneous systems under gravity.

Although we have restricted our study to the case of binary mixtures,
the theory is valid for multicomponent systems, as sedimentation paths
remain lines in the space spanned by all chemical potentials,
cf.~Eq.~(\ref{EQlocalchemichalPotential}). Our theory can also be used
to describe the sedimentation of mixtures with non-colloidal
components. An example are mixtures of platelets and non-adsorbing
polymers where the gravitational length of the polymers is much higher
than that of the colloids. We have checked that the results from our
approach, using the free-volume theory of Ref.~\cite{zhang:9947} to
describe the bulk phase diagram of the mixture, match those of the
effective one-component approach by Wensink {\it et.~al}
\cite{0295-5075-66-1-125} and the experiments by van der Kooij {\it
  et.~al} \cite{PhysRevE.62.5397}. Our theory also applies to
molecular mixtures, provided that the bare rather than the buoyant
masses are used.

\section{Acknowledgements}
We thank Thomas Fischer for useful comments on the manuscript.


\end{document}